\documentclass[1p]{elsarticle}

\usepackage{epsfig}
\usepackage[T1]{fontenc}
\usepackage{afterpage}
\usepackage{amsfonts}
\usepackage{amsmath}
\usepackage{amssymb}
\usepackage{verbatim}
\usepackage{latexsym}
\usepackage{subfig}
\usepackage{slashbox}
\usepackage{multirow}
\usepackage{lscape}
\usepackage{pstricks}
\usepackage{colortbl}
\usepackage{latexsym}
\usepackage{amsthm}

\newcommand{\ep}{\varepsilon}
\newcommand{\pe}{\hspace*{\fill} $\Box$\\}

\newcommand{\prob}{\mbox{Prob\, }}
\newcommand{\bof}[1]{\textbf{#1}}

\newdefinition{definition}{Definition}
\newtheorem{theorem}{Theorem}

\newtheorem{lemma}[theorem]{Lemma}

\begin{document}

\title{The Impossibility of Non-Signaling Privacy Amplification}

\author[CS]{Esther H\"anggi}
\author[PHYS]{Renato Renner} 
\author[CS]{Stefan Wolf}

\address[CS]{
Computer Science Department, ETH Zurich, CH-8092 Zurich, Switzerland.\\
\{esther.haenggi, wolf\}@inf.ethz.ch}
\address[PHYS]{
Institute for Theoretical Physics, ETH Zurich, CH-8093 Zurich, Switzerland
\\ renner@phys.ethz.ch}

\begin{abstract}
Barrett, Hardy, and Kent have shown in 2005 that protocols for quantum key agreement
exist the security of which can be proven under the assumption that quantum \emph{or} relativity 
theory is correct. More precisely, this is based on the non-local behavior of certain 
quantum systems, combined with the non-signaling postulate from relativity. An advantage 
is that the resulting security is independent of what (quantum) systems the legitimate parties'
devices operate on: they do not have to be trusted. Unfortunately, the protocol proposed 
by Barrett {\em et\ al.} cannot tolerate any errors caused 
by noise in the quantum channel. Furthermore, even in the error-free case it is  inefficient: its communication complexity is $\Theta(1/\ep)$ 
when forcing the attacker's information below $\ep$, even if only a single key bit is generated. 
Potentially,  the problem can be solved 
by {\em privacy amplification\/} of relativistic --- or non-signaling --- secrecy. We show, however, that
such privacy amplification is impossible  with respect to the most important form
of non-local behavior, and application of arbitrary hash functions. 
\end{abstract}


\maketitle \thispagestyle{empty}

\section{Introduction, Motivation, and Our Contribution}

\subsection{What is Relativistic Cryptography?}

The security of {\em relativistic cryptography\/} can be proven under the sole 
assumption that the non-signaling postulate of relativity theory is correct. The latter 
states that information transmission faster than at the speed of light is impossible. 
The basic idea, as proposed by Barrett, Hardy, and Kent~\cite{kent}, is as follows: 
By communication over a quantum channel, two parties,  Alice and Bob, generate some 
shared entangled quantum state. They then can carry out measurements and use an authentic classical 
channel to determine the resulting correlation of their respective data. 

So far, 
this is  entanglement-based quantum cryptography as proposed by Ekert~\cite{ekert}\footnote{Interestingly,
the title of Ekert's celebrated article, ``Quantum cryptography based on Bell's theorem,'' suits much 
more precisely --- and might have anticipated in some way --- the idea of relativistic 
cryptography based on non-local correlations: Here, the security proof is directly
based on Bell's theorem, which is not the case for Ekert's protocol.} some years after 
the first quantum key distribution protocol, proposed by Bennett and Brassard~\cite{bb84} that is not 
based on entanglement at all. 
Let us quickly follow Ekert's path:
From the correlations, 
they conclude on error rates and adversarial information and generate a key, the security 
of which can be proven based on the assumption that quantum physics with all its Hilbert-space
formalism is correct~\cite{renatodiss}. An additional assumption that usually has to be made
is that the devices operate on specified quantum systems of given dimension (e.g., single 
polarized photons); the security is lost when the actual systems are different (e.g., pairs of photons). 
The question of device-independent security has been raised already in~\cite{abgs}. 
It was shown that under certain restrictions on the type of possible attacks, namely to 
so-called collective, i.e., {\em i.i.d.}, attacks, it can be achievable at the price of a lower 
key-generation rate.

Let us now turn back to
 {\em relativistic cryptography\/}:
Here, Alice and Bob carry out measurements on their respective systems in a 
space-like separated fashion (to exclude signaling), and this will allow them to
conclude privacy {\em directly\/} from the correlations 
of their resulting  data. The proofs then hold for whatever quantum 
systems the devices operate on; no Hilbert space formalism is used, only classical 
information theory. Actually, the assumption is not even 
necessary that the possibilities of what an adversary can do is limited by  quantum physics. Quantum physics guarantees the protocol to work, i.e., establishes the expected correlations,
the occurrence of which can be verified, 
{\em but the security  is completely independent of quantum physics}. An interesting
consequence is that protocols can be given which are secure if {\em either\/} quantum physics {\em or\/} 
relativity (or both, of course) is correct. 

How can it be possible to derive secrecy from correlations alone? 
In quantum physics, this effect is well-known: Quantum correlations, called {\em entanglement}, 
are monogamous to some extent~\cite{terhal}.
If Alice and Bob are maximally entangled, then Eve 
must be out of the picture. But classically, we do  not know such an 
effect: If Alice and Bob have highly correlated bits, Eve can nevertheless know them. 
The point is that we have to look at correlations of {\em systems}, i.e., bipartite input-output behaviors. 

John Bell has proven in 1964~\cite{bellInequality} that entangled quantum states can display {\em non-local
correlations\/} under measurements. More precisely, the system consists of the choice of the 
particular measurement to be carried out --- the inputs --- and the corresponding outcomes --- the 
outputs. Bell's work was a reply to Einstein, Podolsky, and Rosen's claim~\cite{epr}
 that quantum
physics was incomplete and should be augmented by classical variables determining the behavior 
of every system under any possible measurement. Bell proved that such a thing is impossible: these
variables do not exist. This is what can be exploited cryptographically: If they 
do not exist, then no adversary can have known them before a measurement was carried out.

We explain this 
in more detail and start with a closer look at  systems and correlations.

\subsection{Systems, Correlations, and Non-Locality}

\begin{figure}[h]
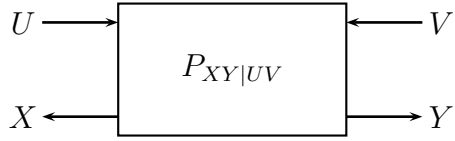

\begin{center}
\pspicture*[](-3.5,-0.2)(3.5,1.95)
\pspolygon[](-1.5,0)(1.5,0)(1.5,1.75)(-1.5,1.75)
\rput[b]{0}(0,0.65){\large{$P_{XY|UV}$}}
\psline[linewidth=1pt]{->}(-2.5,1.5)(-1.5,1.5)
\psline[linewidth=1pt]{<-}(-2.5,0.25)(-1.5,0.25)
\psline[linewidth=1pt]{->}(2.5,1.5)(1.5,1.5)
\psline[linewidth=1pt]{<-}(2.5,0.25)(1.5,0.25)
\rput[b]{0}(-2.75,1.35){\large{$U$}}
\rput[b]{0}(2.75,1.35){\large{$V$}}
\rput[b]{0}(-2.75,0.1){\large{$X$}}
\rput[b]{0}(2.75,0.1){\large{$Y$}}
\endpspicture
\end{center}
\caption{A two-partite system.}
\label{boexli}
\end{figure}

In order to explain the essence of non-locality, we introduce the notion of {\em two-partite 
systems}, defined by their joint input-output behavior $P_{XY|UV}$ (see Figure~\ref{boexli}). 
We classify systems by the correlation they introduce and by the resource required 
to explain the joint behavior of its parts. 

\begin{definition}
{\rm 
\label{localDef}
A {\em system} is a bi- (or more-) partite conditional probability distribution $P_{XY|UV}$. 
A system $P_{XY|UV}$ is {\em independent} if 
$
P_{XY|UV}=P_{X|U}\cdot P_{Y|V}
$.
It is {\em local\/} if 
$
P_{XY|UV}=\sum_{i=1}^n{w_i P_{X|U}^i P_{Y|V}^i}
$
holds for some weights $w_i\geq 0$ and conditional distributions $P_{X|U}^i$ and $P_{Y|V}^i$, $i=1,\ldots,n$. 
A system is {\em signaling\/} if it allows for message transmission, i.e., there exist $P_U$ and $P_V$ such that
$
I(X;V|U)>0
$ or 
$
I(Y;U|V)>0
$
\footnote{
Alternatively, signaling systems can be defined as systems which are \emph{not} non-signaling; a {\em non-signaling}
system being one for which 
\begin{eqnarray}
\nonumber \sum_xP_{XY|UV}(x,y,u,v)&=&\sum_xP_{XY|UV}(x,y,u',v)\ \text{for all}\ y,v,u,u'\ ,\\
\nonumber \sum_yP_{XY|UV}(x,y,u,v)&=&\sum_yP_{XY|UV}(x,y,u,v')\ \text{for all}\ x,u,v,v'
\end{eqnarray}
holds. The two definitions are equivalent. 
}. We call a non-signaling system a \emph{box}.
}
\end{definition}

In terms of {\em classical\/} resources required to establish them, these categories correspond to no resources 
at all, shared information, and message transmission, respectively.
Of interest for us will be systems that are {\em neither local nor signaling}, i.e., non-local boxes. Communication is required to explain 
their behavior classically, but for some of them, distributed {\em quantum\/} information is sufficient. Note that because
they are non-signaling, this does not contradict relativity. 
We give an alternative characterization of locality. 

\begin{lemma}\label{realism}
For any system $P_{XY|UV}$, where ${\cal U}$ and ${\cal V}$ are the ranges of $U$ and $V$, respectively, 
the following  conditions are equivalent:
\begin{enumerate}
\item
$P_{XY|UV}$ is local, 
\item
there exist random variables $X_u$ ($u\in{\cal U}$) and $Y_v$  ($v\in{\cal V}$)
with a joint distribution that is such that the marginals satisfy 
$
P_{X_uY_v}=P_{XY|U=u,V=v}
$.
\end{enumerate}
\end{lemma}
\begin{proof}
Assume first that $P_{XY|UV}$ is local, i.e.,
$
P_{XY|UV}=\sum{w_i P_{X|U}^i P_{Y|V}^i}
$.
For ${\cal U}=\{u_1,u_2,\ldots,u_m\}$ and ${\cal V}=\{v_1,v_2,\ldots,v_n\}$, define
\[
P_{X_{u_1}\cdots X_{u_m}Y_{v_1}\cdots Y_{v_n}}(x_1,\ldots,x_m,y_1,\ldots,y_n):=
\]
\[\sum{w_i P_{X|U=u_1}^i(x_1)\cdots P_{X|U=u_m}^i(x_m)
\cdot P_{Y|V=v_1}^i(y_1)\cdots P_{Y|V=v_n}^i(y_n)}\ .
\]
This distribution has the desired property.

To see the reverse direction, let $X_{u_1}\cdots X_{u_m}Y_{v_1}\cdots Y_{v_n}$ be the shared randomness~$w$. 
\end{proof}

\begin{figure}[h]
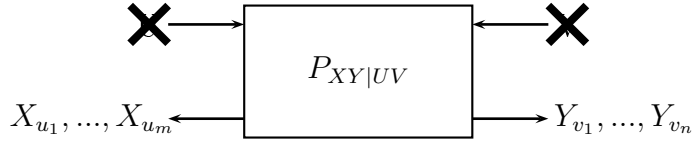

\begin{center}
\pspicture*[](-5,-0.2)(5,1.95)
\pspolygon[](-1.5,0)(1.5,0)(1.5,1.75)(-1.5,1.75)
\rput[b]{0}(0,0.65){\large{$P_{XY|UV}$}}
\rput[b]{0}(-2.75,1.35){\large{U}}
\rput[b]{0}(2.75,1.35){\large{V}}
\psline[linewidth=1pt]{->}(-2.5,1.5)(-1.5,1.5)
\psline[linewidth=1pt]{<-}(-2.5,0.25)(-1.5,0.25)
\psline[linewidth=1pt]{->}(2.5,1.5)(1.5,1.5)
\psline[linewidth=1pt]{<-}(2.5,0.25)(1.5,0.25)
\psline[linewidth=3pt]{-}(-3,1.75)(-2.5,1.25)
\psline[linewidth=3pt]{-}(-2.5,1.75)(-3,1.25)
\psline[linewidth=3pt]{-}(3,1.75)(2.5,1.25)
\psline[linewidth=3pt]{-}(2.5,1.75)(3,1.25)
\rput[b]{0}(-3.5,0.05){\large{$X_{u_1},...,X_{u_m}$}}
\rput[b]{0}(3.5,0.05){\large{$Y_{v_1},...,Y_{v_n}$}}
\endpspicture
\end{center}
\caption{Locality is realism.}
\label{realb}
\end{figure}

Intuitively speaking, we can simply forget about the inputs, and all the alternative 
outputs can be put under the roof of a single joint distribution (see Figure~\ref{realb}).

Lemma~\ref{realism} connects {\em locality\/} with so-called
{\em realism\/}: All the outputs to the alternative inputs ``co-exist'' --- and can, 
hence, be pre-selected --- in a consistent way. In other words, non-locality necessarily means 
that certain data do {\em not  exist\/} before an input is provided on the respective side. 

\subsection{Non-Locality Implies Secrecy}\label{nis}

In order to explain this more explicitly, let us consider a specific example of a system.

\begin{definition}{\bf \cite{pr}}
{\rm 
A {\em Popescu-Rohrlich box\/} (or {\em PR box\/} for short) is the following two-partite 
system $P_{XY|UV}$: The random variable $X$  is a random bit, given the pair $(U,V)$, 
and we have 
\begin{equation}\label{nlbed}
\prob[X\oplus Y=U \cdot V]=1\ .
\end{equation}
}
\end{definition}

Bell's theorem states that this system is indeed non-local. More precisely, 
any system that behaves like a PR box with probability superior to $75\%$ is. Interestingly, the probabilities coming from measurements on bipartite entangled quantum states --- considering the choice of a measurement as the input and the measurement result as the output ---  can achieve roughly $85\%$. 

\begin{theorem}{\bf (John Bell, 1964~\cite{bellInequality}.)}
Any system that behaves like a PR box with probability $>75\%$ for random inputs is non-local.
\end{theorem}

\textit{Proof sketch. }
Lemma~\ref{realism} states that a system is local only if the alternative outputs (i.e., outputs to 
alternative inputs) consistently co-exist. 
In the case of the PR box, this corresponds to a joint distribution of four bits 
$
P_{X_0X_1Y_0Y_1}
$
such that 
$
\prob[X_0=Y_0]=
\prob[X_0=Y_1]=
\prob[X_1=Y_0]=1
$
and
$
\prob[X_1\ne Y_1]=1
$
hold. 
These conditions are contradictory: Only three out of the four can be 
satisfied at a time. \pe 

Note that although in terms of classical resources, the behavior of a PR box can be 
explained by message transmission only, the system is actually non-signaling: $X$ and $Y$
separately are perfectly random bits and independent of the input pair. On the other hand, a system 
 $P_{XY|UV}$ (where all variables are bits) satisfying (\ref{nlbed}) is 
non-signaling {\em only\/} if the outputs are completely unbiased, given the input pair, i.e.,
$
P_{X|U=u,V=v}(0)=
P_{Y|U=u,V=v}(0)=1/2
$.
In other words, the output bit cannot be pre-determined, not even slightly biased.
The outputs are, hence, perfectly random and the randomness must have been generated {\em after\/} 
input reception. 
This is what we can make use of for key agreement: Assume that Alice and Bob share any kind of
 physical system, carry out space-like separated measurements (hereby excluding message transmission), 
and measure data having the statistics of a PR box. (In order to test this, they exchange all 
the input bits and some randomly chosen outputs.) The resulting data are then perfectly secret bits, 
because even conditioned on an adversary's complete information, the correlation between Alice and 
Bob must be non-signaling!

Unfortunately, however, perfect PR boxes do not exist in nature: Quantum physics is non-local, but 
not maximally\footnote{It is a fundamental question, studied by many researchers, why this is the case.
Is there a classical significance to the $85\%$-bound?}. Can we still obtain virtually secret bits from 
weaker, quantum-physically achievable, non-locality? Barrett, Hardy, and Kent~\cite{kent} have shown that 
the answer is {\em yes\/}; but their protocol is inefficient: In order to reduce the probability that
the adversary learns a generated bit shared by Alice and Bob below $\ep$, they have to communicate 
$\Theta(1/\ep)$ Qbits. Barrett {\em et\ al.}'s protocol and its analysis are based on a type of 
non-locality different from the one modeled by the PR box --- the latter is typically referred to as
CHSH~\cite{CHSH} non-locality. 

Masanes and Winter~\cite{mw} proposed to use a number of $85\%$-approximations to the PR box 
(this is achievable with so-called {\em singlets}, i.e., maximally entangled Qbit pairs)\footnote{The analysis of privacy amplification given in the original paper~\cite{mw} led to a result which seems to be in contradiction to our claim that privacy amplification is impossible. In the mean-time it has been realized that the security definition used in~\cite{mw} is incomplete and would only be sufficient if the adversary had to measure before the hash function becomes public (see~\cite{mrwbc} for a revised version).}. Indeed, 
any, even weak, non-locality implies {\em some\/} secrecy, but no perfect secrecy in general. In 
order to illustrate this, consider a system approximating a PR box with probability $1-\ep$ for all 
inputs. More precisely, we have 
\begin{equation}\label{nleps}
\prob[X\oplus Y=U\cdot V|U=u,V=v ]=1-\ep
\end{equation}
for all $(u,v)\in\{0,1\}^2$. Then, what is the maximal possible bias 
\[
p:=\prob[{X=0}|{U=0},{V=0}]
\]
such that the system is non-signaling?
\[
\begin{array}{|c||c|c||c|}
\hline
u & P_{X|U=u,V=v}(0) &  P_{Y|U=u,V=v}(0) & v \\
\hline
\hline
0 & p & p-\ep & 0\\\hline
0 & p & p-\ep & 1\\\hline
1 & p-2\ep & p-\ep & 0\\\hline
1 & p-2\ep & p-\ep & 1\\
\hline
\end{array}
\]
We explain the table: Because of (\ref{nleps}), the bias of $Y$, given $U=V=0$, must be at least $p-\ep$. Because 
of non-signaling, $X$'s bias must be $p$ as well when $V=1$, and so on. Finally, condition (\ref{nleps}) 
for $U=V=1$ implies 
$
p-\ep-(1-(p-2\ep))\leq \ep
$,
hence, 
$
p\leq 1/2+2\ep
$.
For any $\ep<1/4$, this is a non-trivial bound. (This reflects the fact that $\ep=1/4$ is 
the ``local limit,'' as we have seen in the proof of Bell's theorem.) If we apply this, conditioned
on Eve's knowledge, we obtain a lower bound on her uncertainty which is the better the stronger 
the non-locality is. (A special case is what we have seen above already: {\em maximal\/} CHSH non-locality
leads to {\em perfect\/} secrecy.)

In this paper we consider 
privacy amplification applied to the outputs of non-local boxes. Privacy amplification is a concept well-known from classical~\cite{bbr},\, \cite{ill},\, \cite{bbcm}
and quantum~\cite{koenigrenner} cryptography, and means transforming a weakly secret string
into a highly secret key by hashing. 
Because the security of privacy amplification ultimately depends on 
the abilities of an adversary, security in the context of an adversary 
governed by quantum mechanics~\cite{renatodiss} does not necessarily imply security 
in the context of an adversary only restricted by the non-signaling 
condition (modeled by the boxes introduced above). In this latter 
context, security is only known to hold under the additional 
assumption that the adversary can only attack each of the boxes 
separately~\cite{AcinMassarPironio},\, \cite{SGBMPA},\, \cite{AcinGisinMasanes}.  In general, however, an adversary may of course attack 
all of them jointly (corresponding to a coherent attack\footnote{In quantum mechanics, three types of attacks --- individual, collective and coherent attacks --- are generally considered~\cite{BihamMor},\, \cite{BihamMor2},\, \cite{BBBGM}. In an individual attack, the eavesdropper attacks and measures each system identically and independently; in a collective attack the adversary still attacks each system identically and independently, but can make a joint measurement; finally the strongest attack is a coherent attack, where no restrictions apply.}).
Such a more general scenario has been considered by Masanes~\cite{lluis} (see also~\cite{mrwbc}), where the non-signaling postulate is imposed not only between the different parties, but also between subsystems held by one party. In this case, privacy amplification is indeed possible, but this solution requires $n$ devices which are space-like separated and is therefore not practical. 
We consider the general situation where only a space-like separation between Alice and Bob is imposed and Eve can make arbitrary attacks.

\subsection{Our Result: Amplification of Relativistic Privacy is Impossible}

We state our main result informally. We look at the following scenario: in a first phase Alice and Bob have access to $n$ realizations of a $(1-\ep)$-approximation of a PR box. 
In a second phase, they can communicate
over an authentic but public classical channel, and apply arbitrary one-bit-output functions 
to their data set. We then show that, for any such function, they can only reduce the information that a general non-signaling adversary 
can have about the bit, when compared to the raw bits output by the boxes, by at most a factor $4$. 
In other words, privacy amplification by hashing is impossible in the relativistic-crypto\-graphy setting.

\subsection{Outline}

The rest of our paper is organized as follows. In Section~2, we describe the general set of possible 
strategies of a non-signaling adversary. In Section~3, we illustrate the power of a non-signaling adversary
by showing that the XOR of $n$ boxes' output bits 
is not more secure against a non-signaling attack than a single bit --- quite the opposite, in fact. We also describe one
concrete (good) adversarial strategy, which allows the adversary to obtain high information about the key bit. 
This is a special case of our general privacy-amplification no-go result 
stated in Section~4, which shows 
that privacy amplification by 
\emph{any} hash function is impossible.

\section{Preliminaries }
\label{sec:problem}

\subsection{Purification of Bipartite Non-Signaling Systems\protect\footnote{In quantum mechanics, the purification of a bipartite state is the extension of the state to a third party such that the overall state is pure.} }

Assume Alice and Bob share a box. 
When we take the adversary into account, we get a three-partite scenario. The goal of this section is to reduce this tripartite scenario to a bipartite one: 
Given the box Alice and Bob share, what is the most general extension to a third party? This third party will take the role of the adversary. \\

According
to the non-signaling assumption, even the three-partite scenario including the eavesdropper must not allow for signaling.
For instance, we have\footnote{For the sake of simplicity, we drop indices that name the 
random variables when they are obvious.} $P(xy|uvw)=P(xy|uv)$ and $P(x|u)=P(x|uv)$, and so on. 

\begin{figure}[h]
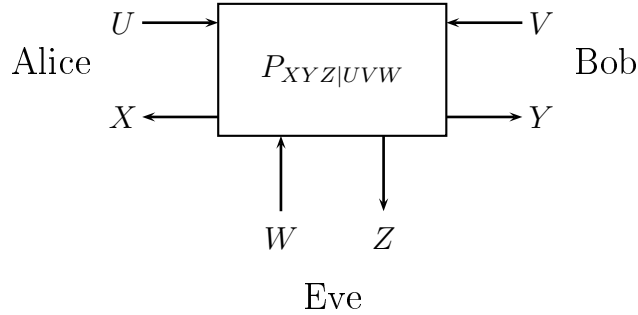

\begin{center}
\pspicture*[](-5,-2.3)(5,1.95)
\pspolygon[](-1.5,0)(1.5,0)(1.5,1.75)(-1.5,1.75)
\rput[b]{0}(0,0.65){\large{$P_{XYZ|UVW}$}}
\psline[linewidth=1pt]{->}(-2.5,1.5)(-1.5,1.5)
\psline[linewidth=1pt]{<-}(-2.5,0.25)(-1.5,0.25)
\psline[linewidth=1pt]{->}(2.5,1.5)(1.5,1.5)
\psline[linewidth=1pt]{<-}(2.5,0.25)(1.5,0.25)
\rput[b]{0}(-2.75,1.35){\large{$U$}}
\rput[b]{0}(2.75,1.35){\large{$V$}}
\rput[b]{0}(-2.75,0.1){\large{$X$}}
\rput[b]{0}(2.75,0.1){\large{$Y$}}
\psline[linewidth=1pt]{->}(-0.675,-1)(-0.675,0)
\psline[linewidth=1pt]{<-}(0.675,-1)(0.675,0)
\rput[b]{0}(-0.6755,-1.5){\large{$W$}}
\rput[b]{0}(0.675,-1.5){\large{$Z$}}
\rput[b]{0}(-3.7,0.82){\Large{Alice}}
\rput[b]{0}(3.6,0.82){\Large{Bob}}
\rput[b]{0}(0,-2.3){\Large{Eve}}
\endpspicture
\end{center}
\caption{The three-partite scenario including the eavesdropper.}
\label{tripartite-situation}
\end{figure}

Note that because of the non-signaling property, the {\em marginal box\/} (in the very same sense
as a marginal probability distribution) of Alice and Bob is well-defined: Their input-output statistics 
do not  depend on what the adversary does, i.e., inputs. 

 The marginal system $P_{XY|UV}$ corresponds to a box. We will use the notation $X$ ($Y$, $U$, $V$) for binary random variables, boldface letters $\bof{X}$ for $n$-bit random variables and $X_i$ for the $i$'th random bit. The values that the random variable take will be denoted by lower-case letters. 
Note that considering only a single box includes the case where Alice and Bob share many boxes, because these can be seen as one box as long as Alice and Bob each give all their inputs simultaneously\footnote{More precisely, Alice gives all her inputs at a given point in space-time and likewise for Bob}. Eve can then attack all these boxes at once through a single input/output interface. 
This is analoguos to an Eve being capable of doing collective attacks in quantum key distribution. Eve's random variables $W$ and $Z$ can have any range. 
Eve's input $w$ corresponds to the choice of her strategy. 
In general, we must assume that Eve can delay her choice until all other parameters are known.
In an informal way, we can write Eve's possibilities as
\begin{eqnarray}
\nonumber \begin{array}{|cc|}\hline
\phantom{_{z_0}}&  \\
\phantom{_{z_0}} A &   B \phantom{_{z_0}}  \\ 
&   \phantom{B}\phantom{_{z_0}} \\\hline
\end{array}
&=&
p({z_0}|w)\cdot
\begin{array}{|cc|}\hline
\phantom{_{z_0}}&  \\
\phantom{_{z_0}} A &   B \phantom{_{z_0}}  \\ 
&  \phantom{B}_{z_0} \\\hline
\end{array}+
p({z_1}|w)\cdot
\begin{array}{|cc|}\hline
\phantom{_{z_0}}&  \\
\phantom{_{z_0}} A &   B \phantom{_{z_0}}  \\ 
&   \phantom{B}_{z_1} \\\hline
\end{array}
+ \ldots \\
\nonumber &=&
p({z'_0}|{w'})\cdot
\begin{array}{|cc|}\hline
\phantom{_{z_0}}&  \\
\phantom{_{z_0}} A &   B \phantom{_{z_0}}  \\ 
&   \phantom{B}_{z'_0} \\\hline
\end{array}
+
p({z'_1}|{w'})\cdot
\begin{array}{|cc|}\hline
\phantom{_{z_0}}&  \\
\phantom{_{z_0}} A &   B \phantom{_{z_0}}  \\ 
&   \phantom{B}_{z'_1} \\\hline
\end{array}
 + \ldots 
\end{eqnarray}

Eve's  strategy $w$ corresponds to a  decomposition of Alice's and Bob's box. Her measurement result $z$ tells her which part of the decomposition occurred. 
\begin{definition}
 {\rm 
A \emph{ box partition}\footnote{This is analogous to quantum mechanics, where bipartite states are described by density operators $\rho_{AB}$ and where any measurement on a purifying system corresponds to a partition of the form $\rho_{AB}=\sum_z p^z \rho_{AB}^z$, where $\rho_{AB}^z$ is the state conditioned on the measurement outcome $z$.} of a given box $P_{\bof{XY}|\bof{UV}}$ is a family of pairs ($p^z$,$P^z_{\bof{XY}|\bof{UV}}$), where $p^z$ is a weight and $P^z_{\bof{XY}|\bof{UV}}$ is a box, such that
 \begin{eqnarray}
 \label{validstrategy} 
P_{\bof{XY}|\bof{UV}}&=&\sum_z p^z\cdot P^z_{\bof{XY}|\bof{UV}}\ . 
\end{eqnarray}
}
\end{definition}
Because of the non-signaling condition, the marginal distribution of Alice and Bob must be the same,  
no matter  Eve's strategy. The fact that any two parties cannot signal to the third party (for example, Alice and Eve together cannot signal to Bob) implies that every box defined by the behavior of the box shared by Alice and Bob conditioned on an outcome $z$, $P_{\bof{XY}|\bof{UV},W=w,Z=z}$, must be  non-signaling. This is, therefore, the most general way to describe a valid strategy of Eve as stated by the following lemmas.
\begin{lemma}
For any given tripartite box $P_{\bof{XY}Z|\bof{UV}W}$ any input $w$ induces a box partition parametrized by $z$: $p^z:=p(z|w)$, $P^z_{\bof{XY}|\bof{UV}}:=P_{\bof{XY}|\bof{UV},Z=z,W=w}$.
\end{lemma}
\begin{lemma}
Given a bipartite box $P_{\bof{XY}|\bof{UV}}$ let $\mathcal{W}$ be a set of box partitions \[w=\{(p^z,P^z_{\bof{XY}|\bof{UV}})\}_z\ .\] Then the tripartite box, where the input of the third part is $w\in\mathcal{W}$, defined by \[P_{\bof{XY}Z|\bof{UV},W=w}(z):=p^z\cdot P^z_{\bof{XY}|\bof{UV}}\] is non-signaling and has marginal box $P_{\bof{XY}|\bof{UV}}$.
\end{lemma}
From now on, $w$ will stand for a certain box partition, i.e., an adversarial strategy. 
We are interested in the question: Which type of box, given outcome $z$, can occur with what probability $p^z$? Let us show that in order to answer this question, it is enough to look at the case where Eve only has {\em two\/} possible measurement outcomes $z_0$ and $z_1$, as all other outcomes can be wrapped into a single one. The reason is that the space of non-signaling boxes is convex. We have
\begin{equation}
\nonumber
\begin{array}{|cc|}\hline
\phantom{_{z_0}}&  \\
\phantom{_{z_0}} A &   B \phantom{_{z_0}}  \\ 
&   \phantom{B}\phantom{_{z_0}} \\\hline
\end{array}
=
p^{z_0}\cdot
\begin{array}{|cc|}\hline
\phantom{_{z_0}}&  \\
\phantom{_{z_0}} A &   B \phantom{_{z_0}}  \\ 
&   \phantom{B}_{z_0} \\\hline
\end{array}
+\underbrace{
p^{z'_1}\cdot
\begin{array}{|cc|}\hline
\phantom{_{z_0}}&  \\
\phantom{_{z_0}} A &   B \phantom{_{z_0}}  \\ 
&   \phantom{B}_{z'_1} \\\hline
\end{array}
+ 
\cdots 
+
p^{z'_m}\cdot
\begin{array}{|cc|}\hline
\phantom{_{z_0}}&  \\
\phantom{_{z_0}} A &   B \phantom{_{z_0}}  \\ 
&   \phantom{B}_{z'_1} \\\hline
\end{array}
}_{p^{z_1}\cdot
\begin{array}{|cc|}\hline
\phantom{_{z_0}}&  \\
\phantom{_{z_0}} A &   B \phantom{_{z_0}}  \\ 
&   \phantom{B}_{z_1} \\\hline
\end{array}}
\end{equation}
where $p^{z_1}=p^{z'_1}+\ldots + p^{z'_m}$.

\

\begin{lemma}\label{eihi} 
If $(p^{z_0},P_{\bof{XY}|\bof{UV}}^{z_0})$ is an element of a box partition with $m$ elements, then it is also an element of a box partition with  {\em only two} elements. 
\end{lemma}
\begin{proof}
We define the probability of outcome $z_1$ as $p^{z_1}=p^{z'_1}+\cdots+ p^{z'_m}$ and the {\em box given outcome $z_1$\/} as 
\begin{eqnarray}
\nonumber P_{\bof{XY}|\bof{UV}}^{z_1}:=\frac{1}{p^{z_1}}\sum\limits_{i=1}^m p^{z'_i}P_{\bof{XY}|\bof{UV}}^{z'_i}\ .
\end{eqnarray}
The marginal distribution defined in this way is the one expected by Alice and Bob because of
\begin{eqnarray}
\nonumber p^{z_0}P_{\bof{XY}|\bof{UV}}^{z_0}+p^{z_1}P_{\bof{XY}|\bof{UV}}^{z_1} &=&p^{z_0}P_{\bof{XY}|\bof{UV}}^{z_0}+\sum\limits_{i=1}^m p^{z'_i}P_{\bof{XY}|\bof{UV}}^{z'_i}=P_{\bof{XY}|\bof{UV}}\ .
\end{eqnarray}

It remains 
 to show that $P_{\bof{XY}|\bof{UV}}^{z_1}$ is a valid non-signaling probability distribution. The convex combination of several probabilities is again a probability between $0$ and $1$ and the normalization remains because every part is normalized separately. 
The distribution is non-signaling because every single part is (the non-signaling property is linear). 
Therefore, $P_{\bof{XY}|\bof{UV}}^{z_1}$ is a valid non-signaling probability distribution, and the two outcomes $z_0$ and $z_1$ define a box partition. 
\end{proof}

\noindent

Lemma~\ref{eihi} implies that for showing an impossibility result, we can assume Eve's information (the random variable $Z$) to be binary. Moreover, we will now show that it is not necessary to determine both parts of the box partition explicitly, but we can find a condition on the box given outcome $z=0$, which will make sure that there exists a second part, complementing it to a box partition.
\begin{lemma}\label{zweihi}
Given a non-signaling  distribution $P_{\bof{XY}|\bof{UV}}$, there exists a box partition with element $(p,P_{\bof{XY}|\bof{UV}}^{Z=0})$ if and only if for all inputs and outputs $\bof{x},\bof{y},\bof{u},\bof{v}$,
\begin{eqnarray}
\label{zcanoccurwithp}  p\cdot P^{Z=0}(\bof{xy}|\bof{uv}) & \leq P(\bof{xy}|\bof{uv})\ .
\end{eqnarray}
\end{lemma}
\begin{proof}
The convex combination of boxes $P_{\bof{XY}|\bof{UV}}^{Z=z}$ is again a box. To prove that the outcome $z=0$ can occur with probability $p$ it is, therefore, needed to show that there exists another valid  outcome $z=1$ which can occur with probability $1-p$, and that the weighted sum of the two is $P_{\bof{XY}|\bof{UV}}$. If $P_{\bof{XY}|\bof{UV}}^{Z=0}$ is a normalized and non-signaling probability distribution, then so is $P_{\bof{XY}|\bof{UV}}^{Z=1}$, because the sum of the two $P_{\bof{XY}|\bof{UV}}$ is also non-signaling and normalized. Therefore, we only need to verify that all entries of the complementary box $P_{\bof{XY}|\bof{UV}}^{Z=1}$ are between $0$ and $1$.  However, this  box is the difference
\begin{eqnarray}
\nonumber P_{\bof{XY}|\bof{UV}}^{Z=1}=\frac{1}{1-p}(P_{\bof{XY}|\bof{UV}}-p\cdot P_{\bof{XY}|\bof{UV}}^{Z=0})\ .
\end{eqnarray}
Requesting this to be greater or equal to $0$ is equivalent to (\ref{zcanoccurwithp}). We observe that all entries of $P_{\bof{XY}|\bof{UV}}^{Z=1}$ are now trivially 
smaller than or equal to  $1$ because of the normalization: if the sum of positive summands is 1, each of them can be at most 1. 
\end{proof}

\subsection{Description of the Scenario and Security Criteria}

We study the scenario where Alice and Bob share several approximations of PR boxes. Alice and Bob can test the behavior of the boxes, but we assume that this has already been done, and that the boxes behave exactly as expected by Alice and Bob; more precisely, we assume that Alice and Bob share $n$ independent and unbiased PR boxes with error $\ep$, defined below.
\begin{definition}
An \emph{unbiased PR box with error $\ep$} is a system $P_{XY|UV}$, where $X,Y,U,V$ are bits, and for every pair $(U,V)$ $X$ and $Y$ are random bits, and
\begin{equation}
\nonumber \prob[X\oplus Y=U \cdot V]=1-\ep
\end{equation}
(see also (\ref{1box_table})).
\end{definition}
The marginal box as seen by Alice and Bob can, therefore, be expressed as
\begin{eqnarray}
\nonumber P_{\bof{XY}|\bof{UV}}&:=& \prod_{i=1}^n P_{X_iY_i|U_iV_i}\ ,
\end{eqnarray}
where $P_{X_iY_i|U_iV_i}$ is a single unbiased PR box with error $\ep$. 
This assumption only restricts Eve's possibilities as compared to the case when the marginal is not fixed. To create a key, Alice and Bob take a public input and a public hash function $f$ 
and apply $f$ to the outcomes of the boxes. The quality of the resulting key can be measured by the distance from the uniform distribution given the adversary's knowledge. In general, the goal of privacy amplification is to create a highly secure bit-string. However, the non-uniformity of the key is lower-bounded by the non-uniformity of a single bit; for showing the impossibility of privacy amplification, it is, therefore, enough to show that the non-uniformity of a single bit is always high. (Indeed, if Alice and Bob cannot even create a single secure bit, they can surely not create several secure bits.)

Since in the protocol, 
 Bob adjusts his output bit to Alice's 
after the exchange of their inputs,
it is enough for Eve to know the output  of Alice's hashing, $f(\bof{x})$. 
For taking into account the most general non-signaling attack, we must assume that Eve can
adapt her strategy to the choice of $f$ and the inputs. 
In our case, it will in fact be sufficient 
for Eve to choose a strategy with only two outputs, $z=0$ and $z=1$,  each occurring with probability ${1}/{2}$, such that given 
$z=0$, $f(\bof{X})$ is maximally biased towards $0$. 
The knowledge Eve has about the key bit $f(\bof{X})$ can be seen as the non-uniformity of this bit, given her outcome $z$. Obviously, this quantity depends on Eve's strategy (the box partition she uses), Alice's and Bob's inputs and the hash function that is applied to the output bits. 

\begin{definition}
The \emph{non-uniformity} of the bit $f(\bof{X})$ given a box partition $w$ and input $\bof{u},\bof{v}$ is 
\begin{eqnarray}
\nonumber \delta_w^f(\bof{u},\bof{v})=\frac{1}{2} \sum\limits_{z} p^{z}\cdot \left|P_{\bof{XY}|\bof{UV}}^z({f(\bof{x})=0|\bof{uv}})-P_{\bof{XY}|\bof{UV}}^z({f(\bof{x})=1|\bof{uv}})\right|\ .
\end{eqnarray}
\end{definition}
Here $\delta_w^f(\bof{u},\bof{v})=0$ means that the eavesdropper has no knowledge about the bit $f(\bof{X})$; on the other hand, $\delta_w^f(\bof{u},\bof{v})=1/2$ corresponds to complete knowledge. 
Our goal will be to show that the non-uniformity remains high, no matter what function Alice (and Bob) apply to their output bits and how many boxes they share.

\subsection{The Case of a Single Box}

In this section, we will show that the knowledge of a non-signaling adversary about the outcome of a box which is non-local is  limited, i.e., we consider
the case where $P_{XY|UV}$ corresponds to a single unbiased PR box with error $\ep$, and the function that is applied is the identity $f=id$.
This is a more detailed justification for the claim already made in Section~\ref{nis}, which also shows that the bound given there is tight.
The marginal probability distribution as seen by Alice and Bob is given by 
\begin{eqnarray}
\label{1box_unbiased_table}
\begin{array}{c c||c|c||c|c||}
$\backslashbox{V}{U}$& & \multicolumn{2}{c||}{0} & \multicolumn{2}{c||}{1} \\
 & $\backslashbox{Y}{X}$ & 0 & 1 & 0 & 1 \\ \hline\hline
\multirow{2}{*}{0} & 0 & \cellcolor{lightgray}\frac{1}{2}-\frac{\ep}{2} & \cellcolor{lightgray}\frac{\ep}{2} & \cellcolor{gray}\frac{1}{2}-\frac{\ep}{2} & \cellcolor{gray}\frac{\ep}{2} \\ \cline{2-6}
& 1 & \frac{\ep}{2} & \frac{1}{2}-\frac{\ep}{2} & \frac{\ep}{2} & \frac{1}{2}-\frac{\ep}{2} \\ \hline\hline
\multirow{2}{*}{1} & 0 & \frac{1}{2}-\frac{\ep}{2} & \frac{\ep}{2} & \frac{\ep}{2} & \frac{1}{2}-\frac{\ep}{2} \\ \cline{2-6}
& 1 & \frac{\ep}{2} & \frac{1}{2}-\frac{\ep}{2} & \frac{1}{2}-\frac{\ep}{2} & \frac{\ep}{2} \\ \hline\hline
\end{array}
\end{eqnarray}
The criteria that Alice cannot signal to Bob translates in this notation to the requirement that the probabilities in the light gray and dark gray areas are equal (and similarly for the other rows); that Bob cannot signal to Alice is expressed as the same kind of condition on the columns. The normalization criteria is that the probabilities within a double line must sum up to one. \\
As an example, we assume that the input bits of Alice and Bob were $(u,v)=(0,0)$; for symmetry reasons it is clear that Eve has an equivalent strategy also for all other inputs.  As shown in \cite{wolf00}, the strategy giving maximal information about a bit to Eve is to choose a box partition with three outputs $z=\{0,1,\delta\}$ such that if she obtains $z=0$ she knows with certainty that Alice's bit $x$ is $0$ and if she obtains $\delta$ Alice's bit is a random bit. 
\begin{eqnarray}
\nonumber P_{XY|UV}&=&p^{Z=0}P^{Z=0}_{XY|UV}+p^{Z=\delta}P^{Z=\delta}_{XY|UV}+p^{Z=1}P^{Z=1}_{XY|UV}\ ,
\end{eqnarray}
where $P^{Z=0}(x=0|u=0,v=0)=1$ and $P^{Z=1}(x=1|u=0,v=0)=1$. From the fact that the marginal box of Alice and Bob is unbiased,
we can conclude $p^{Z=0}=p^{Z=1}$. 

Using this box partition, the non-uniformity of the bit $X$ is given by $\delta_w^{id}(0,0)=1/2(p^{Z=0}+p^{Z=1})=p^{Z=0}$.
However, because of the convexity of the theory, we can easily define another box partition $w'$ with only two outcomes, $z'=0$ and $z'=1$, which also reaches this maximal non-uniformity 
by equally distributing the second ($\delta$) outcome  into the two others:
\begin{eqnarray}
\nonumber P_{XY|UV}&=&p^{Z'=0}P^{Z'=0}_{XY|UV}+p^{Z'=1}P^{Z'=1}_{XY|UV}\ ,
\end{eqnarray}
where $p^{Z'=0}=p^{Z'=1}={1}/{2}$ and 
\begin{eqnarray}
\nonumber
P^{Z'=0}_{XY|UV}&=&2p^{Z=0}P^{Z=0}_{XY|UV}
+\, (1-2p^{Z=0})P^{Z=\delta}_{XY|UV}\ .
\end{eqnarray}
We, therefore, only have to 
find the box that is maximally biased towards $0$ and which can occur with probability $1/2$. 
According to (\ref{zcanoccurwithp}), this turns into a simple maximization problem under linear constraints. The result can be written as follows (where the different entries of the table are $P^{Z=0}_{XY|UV}$):
\begin{eqnarray}
\label{1box_table}
\begin{array}{c c||c|c||c|c||}
$\backslashbox{V}{U}$& & \multicolumn{2}{c||}{0} & \multicolumn{2}{c||}{1} \\
 & $\backslashbox{Y}{X}$ & 0 & 1 & 0 & 1 \\ \hline\hline
\multirow{2}{*}{0} & 0 & \frac{1}{2}+\ep & 0 & \frac{1}{2} & \ep \\ \cline{2-6}
& 1 & \ep & \frac{1}{2}-2\ep & 0 & \frac{1}{2}-\ep \\ \hline\hline
\multirow{2}{*}{1} & 0 & \frac{1}{2}+\ep & 0 & \ep & \frac{1}{2} \\ \cline{2-6}
& 1 & \ep & \frac{1}{2}-2\ep & \frac{1}{2}-\ep & 0 \\ \hline\hline
\end{array}
\end{eqnarray}
This box has  a bias of $2\ep$ towards $0$: We have $P^{Z'=0}(X=0)=1/2+2\ep$, which means that the non-uniformity of Alice's output bit given box partition $w'$ is bounded by $\delta_{w'}^{id}(0,0)=P^{Z'=0}(X=0)-1/2=2\ep$.
This implies that in the case $\ep=0.25$, Eve can perfectly know Alice's output bit --- which corresponds to our expectation since $\ep=0.25$-boxes can be simulated with a local hidden-variable theory, and Eve could know the hidden variable. Additionally, for any 
non-local  theory (i.e., $\ep<0.25$), the non-signaling condition bounds the knowledge of a potential adversary; with {\em perfect\/} PR boxes, one perfectly secret bit per use is created, the confidentiality of which relies only on the non-signaling condition, as we have explained 
in Section~\ref{nis}.

\section{Impossibility of Privacy Amplification by Linear Hashing}
\label{sec:privamp}

In this section, we  show that privacy amplification by applying a  linear function --- taking the XOR of some subset of the output bits --- is impossible. Moreover, we will show that the more bits we take the XOR of, the more Eve can know. At the same time we try to give a more intuitive explanation of the possibilities Eve has, and explain why the strategy we give is actually a good strategy for Eve. The specific box partition we define here will also be used later in the case of general hash functions.

\subsection{Intuitive Presentation of the Argument}

Let us take as an example the case where Alice and Bob share only two boxes. Then we can define a table with input-output probabilities similar to (\ref{1box_unbiased_table}) for the two boxes $P_{X_1X_2Y_1Y_2|U_1U_2V_1V_2}$. 
 Alice and Bob now have two bits of input and output. As the two boxes as seen by Alice and Bob are independent, the probabilities 
are simply the product of the input-output probabilities of each box. We give a part of this table --- the part for the input 
$(u_1,u_2,v_1,v_2)=(0,0,0,0)$:
\begin{center}
\begin{displaymath}
\begin{array}{cc||c|c|c|c||cccccccccccc}
$\backslashbox{$V_1V_2$}{$U_1U_2$}$ & & \multicolumn{4}{c||}{00} & \multicolumn{4}{c}{}& \multicolumn{4}{c}{} & \multicolumn{4}{c}{} \\
 & $\backslashbox{$Y_1Y_2$}{$X_1X_2$}$ &  \cellcolor{lightgray} 00 & 01 & 10 &  \cellcolor{lightgray} 11 &  \\ \hline \hline
\multirow{8}{*}{00} & \multirow{2}{*}{00} & \cellcolor{lightgray} (\frac{1}{2}-\frac{\ep}{2})\cdot & (\frac{1}{2}-\frac{\ep}{2})\cdot & \frac{\ep}{2}\cdot &  \cellcolor{lightgray}\frac{\ep}{2}\cdot &  \multirow{2}{*}{$\Longleftarrow$ row 1} \\
&  &  \cellcolor{lightgray} (\frac{1}{2}-\frac{\ep}{2}) &  \frac{\ep}{2} & (\frac{1}{2}-\frac{\ep}{2}) &   \cellcolor{lightgray}\frac{\ep}{2}     
\\ \cline{2-18}
& \multirow{2}{*}{01} &  \cellcolor{lightgray} (\frac{1}{2}-\frac{\ep}{2})\cdot & (\frac{1}{2}-\frac{\ep}{2})\cdot & \frac{\ep}{2}\cdot &  \cellcolor{lightgray}\frac{\ep}{2}\cdot &  \\
& &  \cellcolor{lightgray} \frac{\ep}{2} &  (\frac{1}{2}-\frac{\ep}{2}) & \frac{\ep}{2} &  \cellcolor{lightgray} (\frac{1}{2}-\frac{\ep}{2}) &   
\\ \cline{2-18}
& \multirow{2}{*}{10} &  \cellcolor{lightgray} \frac{\ep}{2}\cdot & \frac{\ep}{2} \cdot& (\frac{1}{2}-\frac{\ep}{2})\cdot &  \cellcolor{lightgray}(\frac{1}{2}-\frac{\ep}{2}) \cdot  & \ldots \\
&  &  \cellcolor{lightgray}(\frac{1}{2}-\frac{\ep}{2}) &  \frac{\ep}{2} & (\frac{1}{2}-\frac{\ep}{2}) &   \cellcolor{lightgray}\frac{\ep}{2}     &    
\\ \cline{2-18}
& \multirow{2}{*}{11} &  \cellcolor{lightgray}\frac{\ep}{2}\cdot & \frac{\ep}{2}\cdot & (\frac{1}{2}-\frac{\ep}{2})\cdot & \cellcolor{lightgray} (\frac{1}{2}-\frac{\ep}{2})\cdot   & \\ 
& & \cellcolor{lightgray} \frac{\ep}{2} &  (\frac{1}{2}-\frac{\ep}{2}) & \frac{\ep}{2} &   \cellcolor{lightgray}(\frac{1}{2}-\frac{\ep}{2})  & 
\\ \hline \hline
 &  &  &\vdots  &  &  &   \\
\end{array}
\end{displaymath}
\end{center}
Now, imagine further that Eve learns the input (later, we will see that this is, in fact, not necessary for her strategy to work), and that the function Alice applies is the XOR (which is the only non-trivial linear function). In the table we  mark gray all outcomes for which Alice's final bit is $0$ and as white the ones where she will get $1$. What is Eve's strategy which gives her as much information as possible about Alice's final bit? Defining a strategy means  constructing a decomposition of the double box into two parts corresponding to her two (equally likely) outputs:
\begin{eqnarray}
\nonumber 
\begin{array}{|cc|}\hline
\phantom{_{z_0}}&  \\
\phantom{_{z_0}} A &   B \phantom{_{z_0}}  \\ 
&   \phantom{B}\phantom{_{z_0}} \\\hline
\end{array}
&=&
\frac{1}{2}\cdot
\begin{array}{|cc|}\hline
\phantom{_{z_0}}&  \\
\phantom{_{z_0}} A &   B \phantom{_{z_0}}  \\ 
&  \phantom{B}_{z=0} \\\hline
\end{array}+
\frac{1}{2}\cdot
\begin{array}{|cc|}\hline
\phantom{_{z_0}}&  \\
\phantom{_{z_0}} A &   B \phantom{_{z_0}}  \\ 
&   \phantom{B}_{z=1} \\\hline
\end{array}
\end{eqnarray}
In fact, as we have seen before, we only need to construct one part  $P^{Z=0}_{\bof{XY}|\bof{UV}}$ such that $z=0$ can occur with probability $1/2$ according to (\ref{zcanoccurwithp}); the second part is then automatically defined. As Eve has learned the input $0000$ and knows that the function Alice applies is the XOR, we try to make the box given measurement outcome $z=0$ maximally biased towards $0$, that is, Alice's output bits should be likely to be either $00$ or $11$. In order to construct the conditional box, given measurement outcome $z=0$, we start from the unbiased box as seen by Alice and Bob (given above) and shift around probabilities. More precisely, we try to take as much probability as possible out of the white area and put it into the gray area, therefore, biasing the XOR towards~$0$. However, we have to 
respect a few rules.
\begin{enumerate}
 \item \label{item:probability}All entries must remain \textbf{probabilities} between $0$ and $1$.
 \item The \textbf{normalization} of the probability distribution must remain --- this will not be a problem as we only move around probability weights within the same input, taking them out of one cell and putting them into another.
 \item The \textbf{non-signaling} condition must be satisfied --- this implies that even the input is known, we must be able to define the conditional box, given output $z=0$, for all inputs, and it must be possible to do this in a non-signaling way. We will not worry about this condition too much for the moment, as we will be able to show later that if we proceed as below for all inputs, the box obtained is in fact non-signaling. 
 \item \label{item:compensate} There \textbf{must exist a second measurement outcome} $z=1$ occurring with probability $1/2$, and such that the conditional box, given outcome $z=1$, is also a valid probability distribution. This box, given outcome $z=1$, must be able to compensate for the shifts in probabilities. According to (\ref{zcanoccurwithp}), this means that the entry in every cell must be smaller or equal twice the original entry. 
\end{enumerate}
Rules \ref{item:probability} and \ref{item:compensate} together state that every new entry must be between $0$ and twice the original entry. 

Now, we can proceed row-wise in the picture (a row corresponds to one specific output on Bob's side) and look at the probabilities in the gray and white areas --- i.e., the probability for Alice's final bit to be $0$ or $1$ respectively. In case of the first row (corresponding to Bob's output $\bof{y}=y_1y_2=00$), we see that the probability for Alice to obtain $00$ or $11$ is $(1/2-\ep/{2})^2$ and $(\ep/{2})^2$, respectively, and to obtain $01$ or $10$ is $({1}/{2}-{\ep}/{2})(\ep/{2})$ each. We try to take as much probability as possible out of the white area and put it into the gray area. If the sum of probabilities in the white area is smaller than the one in the gray area, we can take \emph{all} probability weight out of the white cells and distribute it in the gray cells proportionally to the original entry. As the white area is smaller than the gray, this will at most  double the gray entries, and all entries will be within the range allowed by (\ref{zcanoccurwithp}). This is the case here:  $({1}/{2}-{\ep}/{2})^2+({\ep}/2)^2>2({1}/{2}-{\ep}/{2})({\ep}/{2})$, hence, the new entries in the white cells will be $0$, and the new entries in the gray cells will be less than twice the amount that was there before. We will call rows of this type $\bof{y}_>$ (\ref{y>}), and a generalization of this argument will lead to (\ref{px0y>}), (\ref{px1y>}).

The second row  is different: The probabilities in the gray area are lower than the ones in the white area ($2({1}/{2}-{\ep}/{2})({\ep}/{2})<({1}/{2}-{\ep}/{2})^2+({\ep}/{2})^2$), so we cannot shift the entire probability into the gray cells, because this would more than double the entries. The best we can do is to exactly double the entries in the gray region, and take exactly this probability (proportionally) out of the white cells, which means the amount of probability that is shifted is $2({1}/{2}-{\ep}/{2})({\ep}/{2})$. This type of row will be called $\bof{y}_<$ (\ref{y<}), and the (generalized) expression of the new entries is given in (\ref{px0y<}), (\ref{px1y<}).\\

So, whether we look at a row of type $\bof{y}_<$ or $\bof{y}_>$, the amount of probability that is shifted is exactly the probability contained in the area with lower total probability (gray or white). The shifted probability will correspond exactly to the bias the box given outcome $z=0$ has, so we just have to count the lower of the two areas on every row. This is what is said in (\ref{biaswbar}). And this bias is exactly the non-uniformity of the key bit given box partition $\bar{w}$. In our example of two boxes and the XOR, the probability shift (area with the lower probability) happens to be $2({1}/{2}-{\ep}/{2})(\ep/2)$ for every row, and there are four rows, therefore, 
$\delta_{example}^{XOR}(00,00)=4\cdot 2({1}/{2}-{\ep}/{2})({\ep}/{2})=2\ep-2\ep^2$.
Finally, note that if we shift probabilities in this way for all inputs, the resulting box is in fact non-signaling. The fact that Alice cannot signal to Bob is satisfied because we do not shift probabilities between rows. Bob cannot signal to Alice because for every row, the same row (meaning: with the same sequence of probabilities) appears again for another input of Bob (maybe just in a different position) and the same shifts, therefore, occur for all inputs of Bob.

\subsection{A Concrete (Good) Adversarial Strategy}
\label{subsec:wbar}

Now we will define formally the box partition we  described 
informally in the previous section. 
We describe a box partition $\bar{w}$ which contains an element $(1/2,P^{Z=0}_{\bof{XY}|\bof{UV}})$, and which gives a high non-uniformity of the key bit $f(\bof{X})$. Our description will be rather general, such that this box partition can also be used when the function $f$ is not the XOR. 
The probabilities $P^{Z=0}(\bof{x},\bof{y},\bof{u},\bof{v})$ are defined in four cases according to $\bof{x}$, $\bof{y}$ and the properties of the box $P_{\bof{XY}|\bof{UV}}$ (in terms of the intuitive   explanation above: whether we are in a white or gray cell, whether there is more probability in the white or the gray area and according to the original entry in that cell). For simplicity, let us use the following notation:
\begin{eqnarray}
\label{y<}\bof{y}_<&:=& \left\{\bof{y}| \sum\limits_{\bof{x}|f(\bof{x})=0} P({\bof{xy}|\bof{uv}})<\sum\limits_{\bof{x}|f(\bof{x})=1} P({\bof{xy}|\bof{uv}})\right\}\ ,\\
\label{y>}\bof{y}_> &:=& \left\{\bof{y}| \sum\limits_{\bof{x}|f(\bof{x})=0} P({\bof{xy}|\bof{uv}})>\sum\limits_{\bof{x}|f(\bof{x})=1} P({\bof{xy}|\bof{uv}})\right\}\ ,\\
\label{x0}\bof{x}_0 &:=& \{\bof{x}|f(\bof{x})=0\}\ ,\\
\label{x1}\bof{x}_1 &:=& \{\bof{x}|f(\bof{x})=1\}\ .
\end{eqnarray}
Then $P^{Z=0}(\bof{xy}|\bof{uv})$ is defined as follows:\\
For all $\bof{x}\in \bof{x}_0,\bof{y} \in \bof{y}_<:$
\begin{eqnarray}
\label{px0y<} P^{Z=0}(\bof{xy}|\bof{uv})&:=&2\cdot P({\bof{xy}|\bof{uv}})\ .
\end{eqnarray}
For all $\bof{x}\in \bof{x}_1,\bof{y}\in \bof{y}_<:$
\begin{eqnarray}
\label{px1y<} P^{Z=0}(\bof{xy}|\bof{uv}) &:=&\frac{\sum\limits_{\bof{x}|f(\bof{x})=1} P({\bof{xy}|\bof{uv}})-\sum\limits_{\bof{x}|f(\bof{x})=0} P({\bof{xy}|\bof{uv}})}{\sum\limits_{\bof{x}|f(\bof{x})=1} P({\bof{xy}|\bof{uv}})}\cdot P({\bof{xy}|\bof{uv}})\ . 
\end{eqnarray}
For all $\bof{x}\in \bof{x}_0,\bof{y}\in \bof{y}_>:$
\begin{eqnarray}
\label{px0y>} P^{Z=0}(\bof{xy}|\bof{uv}) &:=&\frac{\sum\limits_{\bof{x}|f(\bof{x})=1} P({\bof{xy}|\bof{uv}})+\sum\limits_{\bof{x}|f(\bof{x})=0} P({\bof{xy}|\bof{uv}})}{\sum\limits_{\bof{x}|f(\bof{x})=0} P({\bof{xy}|\bof{uv}})}\cdot P({\bof{xy}|\bof{uv}})\ .
\end{eqnarray}
For all $\bof{x}\in \bof{x}_1,\bof{y}\in \bof{y}_>:$
\begin{eqnarray}
\label{px1y>} P^{Z=0}(\bof{xy}|\bof{uv})&:=&0\ .
\end{eqnarray}

\begin{lemma} 
There exists a box partition with an element $(p^{Z=0}={1}/{2},P^{Z=0}(\bof{xy}|\bof{uv}))$.
\end{lemma}
\begin{proof}\emph{\phantom{A }}\\
\emph{Alice cannot signal to Bob}:\\
For all $\bof{u},\bof{v}$ and $\bof{y}\in \bof{y}_<$:
\begin{eqnarray}
\label{ns-AtoB1}&&\sum\limits_{\bof{x}} P^{Z=0}(\bof{xy}|\bof{uv}) =\sum\limits_{\bof{x}|f(\bof{x})=0}2\cdot P({\bof{xy}|\bof{uv}})+
\\
\nonumber && \sum\limits_{\bof{x}|f(\bof{x})=1}\frac{\sum\limits_{\bof{x'}|f(\bof{x'})=1} P({\bof{xy}|\bof{uv}})-\sum\limits_{\bof{x'}|f(\bof{x'})=0} P({\bof{xy}|\bof{uv}})}{\sum\limits_{\bof{x'}|f(\bof{x'})=1} P({\bof{xy}|\bof{uv}})}
\cdot P({\bof{xy}|\bof{uv}})\\
\nonumber &=& 2 \sum\limits_{\bof{x}|f(\bof{x})=0} P({\bof{xy}|\bof{uv}})+\sum\limits_{\bof{x}|f(\bof{x})=1} P({\bof{xy}|\bof{uv}})
-\sum\limits_{\bof{x}|f(\bof{x})=0} P({\bof{xy}|\bof{uv}})\\
\nonumber&=& \sum\limits_{\bof{x}} P({\bof{xy}|\bof{uv}})=\frac{1}{2^n}\ .
\end{eqnarray}
For all $\bof{u},\bof{v}$ and $\bof{y}\in \bof{y}_>$:
\begin{eqnarray}
\label{ns-AtoB2} &&\sum\limits_{\bof{x}} P^{Z=0}(\bof{xy}|\bof{uv}) = 
\sum\limits_{\bof{x}|f(\bof{x})=1}0 
\\ \nonumber &&
+\sum\limits_{\bof{x}|f(\bof{x})=0}\frac{\sum\limits_{\bof{x'}|f(\bof{x'})=1} P({\bof{xy}|\bof{uv}})+\sum\limits_{\bof{x'}|f(\bof{x'})=0} P({\bof{xy}|\bof{uv}})}{\sum\limits_{\bof{x'}|f(\bof{x'})=0} P({\bof{xy}|\bof{uv}})}
\cdot P({\bof{xy}|\bof{uv}})\\
\nonumber &=&\sum\limits_{\bof{x}|f(\bof{x})=1} P({\bof{xy}|\bof{uv}})+\sum\limits_{\bof{x}|f(\bof{x})=0} P({\bof{xy}|\bof{uv}})=\frac{1}{2^n}\ .
\end{eqnarray}
\\
\emph{Bob cannot signal to Alice}: 
For this, we need the fact that 
\begin{eqnarray}
\nonumber P({\bof{xy}|\bof{uv'}})&=& 
P({\bof{xy'}|\bof{uv}})\ ,
\end{eqnarray}
where the $i$'th bit of $\bof{y'}$ is defined as 
$y'_i:=y_i\oplus u_i\cdot(v'_i-v_i)$ and,
therefore, we have for all $\bof{x},\bof{u},\bof{v}$:
\begin{eqnarray}
\label{ns-BtoA} \sum\limits_{\bof{y}} P^{Z=0}(\bof{xy}|\bof{uv'})&=&\sum\limits_{\bof{y'}} P^{Z=0}(\bof{xy'}|\bof{uv})=\sum\limits_{\bof{y}} P^{Z=0}(\bof{xy}|\bof{uv})\ .
\end{eqnarray}\\
\emph{Normalization}:
This follows directly from (\ref{ns-AtoB1}), (\ref{ns-AtoB2}):
\begin{eqnarray}
\nonumber \sum\limits_{\bof{x},\bof{y}}  P^{Z=0}(\bof{xy}|\bof{uv})&=&\sum\limits_{\bof{y}}\left( \sum\limits_{\bof{x}} P^{Z=0}(\bof{xy}|\bof{uv})\right)=\sum\limits_{\bof{y}}\frac{1}{2^n}=1\ .
\end{eqnarray}\\
\emph{$p^{Z=0}={1}/{2}$}: 
For the case $p={1}/{2}$  (\ref{zcanoccurwithp}) translates to $P^{Z=0}(\bof{xy}|\bof{uv})\leq 2\cdot P({\bof{xy}|\bof{uv}})$, which is satisfied due to the definition of $P^{Z=0}(\bof{xy}|\bof{uv})$. 
\end{proof}

We can define a complementary box $P^{Z=1}(\bof{xy}|\bof{uv})=2\cdot P({\bof{xy}|\bof{uv}})-P^{Z=0}(\bof{xy}|\bof{uv})$, to give a box partition
\begin{eqnarray}
\nonumber P_{\bof{XY}|\bof{UV}}=\frac{1}{2}P^{Z=0}(\bof{xy}|\bof{uv}) +\frac{1}{2}P^{Z=1}(\bof{xy}|\bof{uv})\ .
\end{eqnarray}
The bias of the bit $f(\bof{X})$ for the box $P^{Z=0}_{\bof{XY}|\bof{UV}}$ and, therefore, also $\delta_{\bar{w}}^f(\bof{u},\bof{v})$ is given by 
\begin{eqnarray}
\label{biaswbar}\delta_{\bar{w}}^f(\bof{u},\bof{v})=\sum\limits_{\bof{y}} \min\left\{\sum\limits_{\bof{x}|f(\bof{x})=0} P({\bof{xy}|\bof{uv}}),\sum\limits_{\bof{x}|f(\bof{x})=1} P({\bof{xy}|\bof{uv}})\right\}\ ,
\end{eqnarray}

\subsection{The Impossibility Result}

We will now show that Alice cannot use any linear function --- XOR of some of her output bits --- to do privacy amplification. There always exists a 
box partition
(namely $\bar{w}$) such that 
the non-uniformity of the key bit given $\bar{w}$ is bigger than $\ep$, the error of the box: $\delta_w^f(\bof{u},\bof{v})\geq \ep$.
Furthermore, taking the XOR of many output bits is actually counter-productive, as the non-uniformity of the key bit grows in the number of bits the XOR is taken of, and in the limit of large $n$, Eve can even have close-to-perfect knowledge about Alice's final bit. 

\begin{lemma}
For all linear hash functions $f:\{0,1\}^n\rightarrow \{0,1\}$, the non-uniformity of the bit $f(\bof{X})$ given box partition $\bar{w}$ is larger than $\ep$: $\delta_{\bar{w}}^f(\bof{u},\bof{v})\geq \ep$.
\end{lemma}
\begin{figure}[ht!]
\centering
\includegraphics[width=7cm]{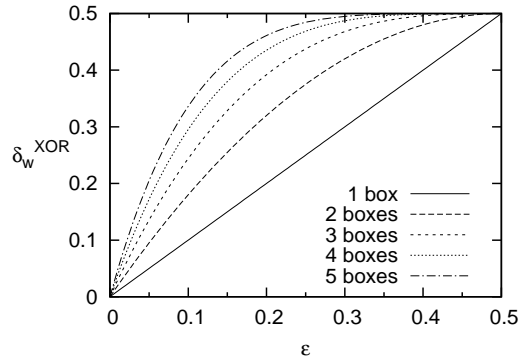}
\caption{The lower bound on the non-uniformity of the final bit (XOR of all outputs) as given by (\ref{eq:biasxor}) as function of number of boxes and error. Note that the non-trivial region of $\ep$ is below $0.25$.}
\end{figure}
\begin{proof}
If the function $f$ is the XOR, we can explicitly determine the non-uniformity of the bit $f(\bof{X})$ given the box partition $\bar{w}$. In fact, we have either 
\begin{eqnarray}
\nonumber \sum\limits_{\bof{x}|\oplus_i x_i=0}P({\bof{xy}|\bof{uv}})&=&\binom{n}{n}\left(\frac{1}{2}-\frac{\ep}{2}\right)^n
+\binom{n}{n-2}\left(\frac{1}{2}-\frac{\ep}{2}\right)^{n-2}\left(\frac{\ep}{2}\right)^{2}\\
\nonumber &&+\binom{n}{n-4}\left(\frac{1}{2}-\frac{\ep}{2}\right)^{n-4}\left(\frac{\ep}{2}\right)^{4}+...\\
\nonumber \sum\limits_{\bof{x}|\oplus_i x_i =1}P({\bof{xy}|\bof{uv}})&=&\binom{n}{n-1}\left(\frac{1}{2}-\frac{\ep}{2}\right)^{n-1}\left(\frac{\ep}{2}\right)^{1}\\
\nonumber&&+
\binom{n}{n-3}\left(\frac{1}{2}-\frac{\ep}{2}\right)^{n-3}\left(\frac{\ep}{2}\right)^{3}
+...
\end{eqnarray}
 or \textit{vice versa}. Therefore, 
\begin{eqnarray}
\nonumber \delta_{\bar{w}}^{XOR}(\bof{u},\bof{v})
&=&\sum\limits_y \min\left\{ \binom{n}{n}\left(\frac{1}{2}-\frac{\ep}{2}\right)^n+
\binom{n}{n-2}\left(\frac{1}{2}-\frac{\ep}{2}\right)^{n-2}\left(\frac{\ep}{2}\right)^{2}+
...,\right.\\
\nonumber&&\left.\binom{n}{n-1}\left(\frac{1}{2}-\frac{\ep}{2}\right)^{n-1}\left(\frac{\ep}{2}\right)^{1}+
\binom{n}{n-3}\left(\frac{1}{2}-\frac{\ep}{2}\right)^{n-3}\left(\frac{\ep}{2}\right)^{3}+
...\right\}\\
\nonumber&=&\sum\limits_y\left( \binom{n}{n-1}\left(\frac{1}{2}-\frac{\ep}{2}\right)^{n-1}\left(\frac{\ep}{2}\right)^{1}
+
\binom{n}{n-3}\left(\frac{1}{2}-\frac{\ep}{2}\right)^{n-3}\left(\frac{\ep}{2}\right)^{3}+... \right)\\
\label{eq:biasxor}&=& \sum\limits_{i=1}^{n/2}\binom{n}{n-(2i-1)}\left(1-\ep\right)^{n-(2i-1)}\ep^{2i-1}\ ,
\end{eqnarray}
which is larger than $\ep$ for all $n>1$, and shows that there exists a constant lower bound on the knowledge Eve can always obtain about the key bit
 by using this strategy. Furthermore, at the limit of large $n$, the non-uniformity of the bit $f(\bof{X})$ tends toward $1/2$ and Eve can have almost perfect knowledge about Alice's output bit, no matter the original error of the box. 
\end{proof}

\section{Impossibility of Privacy Amplification by Any  Hashing}

In this section, we look at  the case where Alice and Bob  apply any function of their choice to their output bits. 
First we will show that by using the box partition $\bar{w}$ defined in Section~\ref{subsec:wbar}, Eve can gain the same knowledge for a given function, no matter what the input of Alice and Bob was. This property will make the argument simpler, as it will be sufficient to look at the non-uniformity of the bit $f(\bof{X})$ for the all-$0$ input. 
\begin{lemma}\label{barka}
The non-uniformity of the bit $f(\bof{X})$ conditioned on the box partition $\bar{w}$ defined in Section~\ref{subsec:wbar} is independent of the values of $\bof{u},\bof{v}$, i.e., $\delta_{\bar{w}}^f(\bof{u},\bof{v})=\delta_{\bar{w}}^f$.
\end{lemma}
\begin{proof}
Let us first express the probability of the output $\bof{x},\bof{y}$, given input $\bof{u},\bof{v}$, as a function of the probabilities given the $0$-input.
\begin{eqnarray}
\nonumber P({\bof{xy}|\bof{uv}})&=&\left(\frac{1}{2}-\frac{\ep}{2}\right)^{(\sum\limits_i 1\oplus x_i \oplus y_i\oplus u_i\cdot v_i)}\cdot \left(\frac{\ep}{2}\right)^{(\sum\limits_i x_i \oplus y_i\oplus u_i\cdot v_i)}\\
\nonumber &=&
P({\bof{xy'}|0...0})\ ,
\end{eqnarray}
where we have again 
defined $y'_i=y_i\oplus u_i\cdot v_i$. Therefore,
\begin{eqnarray}
\nonumber \delta_{\bar{w}}^f(\bof{u},\bof{v}) &=& \sum\limits_{\bof{y}} \min\left\{\sum\limits_{\bof{x}|f(\bof{x})=0} P({\bof{xy}|\bof{uv}}),\sum\limits_{\bof{x}|f(\bof{x})=1} P({\bof{xy}|\bof{uv}})\right\}\\
\nonumber &=&\sum\limits_{\bof{y'}} \min\left\{\sum\limits_{\bof{x}|f(\bof{x})=0} P(\bof{xy'}|0...0),\sum\limits_{\bof{x}|f(\bof{x})=1} P(\bof{xy'}|0...0)\right\}\\
\nonumber &=&\sum\limits_{\bof{y}} \min\left\{\sum\limits_{\bof{x}|f(\bof{x})=0} P({\bof{xy}|0...0}),\sum\limits_{\bof{x}|f(\bof{x})=1} P({\bof{xy}|0...0})\right\}\\
\nonumber  &=& \delta_{\bar{w}}^f(0...0,0...0)\ .
\end{eqnarray}
\end{proof}

Hence, we only have to find a lower bound on the 
non-uniformity 
$
\delta_{\bar{w}}^f(0...0,0...0)
$, 
which we can explicitly write as
\begin{eqnarray}
\nonumber \delta_{\bar{w}}^f(0...0,0...0) &=& \sum\limits_{\bof{y}} \min\left\{\sum\limits_{\bof{x}|f(\bof{x})=0}\left(\frac{1}{2}-\frac{\ep}{2}\right)^{n-d_H(\bof{x},\bof{y})}\cdot \left(\frac{\ep}{2}\right)^{d_H(\bof{x},\bof{y})},\right.\\
\nonumber &&\left.\sum\limits_{\bof{x}|f(\bof{x})=1}\left(\frac{1}{2}-\frac{\ep}{2}\right)^{n-d_H(\bof{x},\bof{y})}\cdot \left(\frac{\ep}{2}\right)^{d_H(\bof{x},\bof{y})}\right\}\ .
\end{eqnarray}
\noindent
Before we continue, we have to introduce some basic facts about correlations:

We will use the following expression for the \emph{distance from uniform} of a random bit $X$: 
\begin{eqnarray}
\nonumber \delta(P_X,P_U)&=&\max(P_X(0),P_X(1))-\frac{1}{2}\ .
\end{eqnarray}
The \emph{correlation $c_{XY}$} between two random bits $X$ and $Y$ is the probability for the two bits to be equal, minus the probability
for the two bits to be different
\begin{eqnarray}
\nonumber c_{XY}&=&P(X=Y)-P(X\neq Y)\ .
\end{eqnarray}
Two equal random bits  have correlation $1$ and are called \emph{completely correlated}, two random bits which are always different have correlation $-1$ and are called \emph{completely anti-correlated}.

Let us further notice here the following: Assume Alice has a random bit-string $\bof{X}$ to which she applies a public function $f$ to obtain a single bit: $f:\bof{X}\to \{0,1\}$. Bob has a random bit-string $\bof{Y}$ which is correlated with $\bof{X}$ and he would like to calculate a bit $Z$ that is highly correlated with $f(\bof{X})$. Then the best achievable correlation is $c_{Zf(\bof{X})}^{\mathrm{opt}}=2\mathbb{E}_{\bof{y}}[\max(P(f(\bof{X})=0|Z=z),P(f(\bof{X})=1|Z=z))]-1$, and it is reached by choosing $Z$ to be $0$ (respectively $1$) if $f(\bof{X})$ is more likely to be $0$ ($1$) given the information $\bof{Y}$.
\begin{definition}
{\rm 
Assume a random variable $\bof{X}$, which is mapped to a bit $f(\bof{X})\in \{0,1\}$, and a random variable $\bof{Y}$ giving some information about the value of $\bof{X}$.  
The \emph{maximum likelihood function g} of $f(\bof{X})$ given $\bof{Y}$ is the function $g:\bof{Y}\rightarrow \{0,1\}$ defined as
\begin{eqnarray}
\nonumber g(\bof{y})&=&
\left\lbrace
{\begin{array}{ll}
0 & \text{if}\ P(f(\bof{X})=0|\bof{Y}=\bof{y})\geq P(f(\bof{X})=1|\bof{Y}=\bof{y})\\
1 & \text{if}\ P(f(\bof{X})=0|\bof{Y}=\bof{y}) < P(f(\bof{X})=1|\bof{Y}=\bof{y})\ .
\end{array}
}
\right.
\end{eqnarray}
}
\end{definition}

With these definitions, we can now show the key lemma for the derivation of our result. It states that Eve can always obtain knowledge proportional to the error in correlation between Alice's and Bob's key bit. 
\begin{lemma}\label{lemma:knowledgereduce}
There exists a box partition $w$ such that $\delta_w^f=1/2-1/2\cdot c_{f(\bof{X})g(\bof{Y})}$, where $f,g:\{0,1\}^n\rightarrow \{0,1\}$, and where $g$ is the maximum likelihood function of $f(\bof{X})$ given $\bof{Y}$.
\end{lemma}
\begin{proof}
It suffices to show that the 
box partition
$\bar{w}$ given in Section~\ref{subsec:wbar} (see (\ref{biaswbar})) reaches this bound:
\begin{small}
\begin{eqnarray}
\nonumber \delta_{\bar{w}}^f 
&=& \frac{1}{2^n}\sum\limits_{\bof{y}} \min\left\{\sum\limits_{\bof{x}|f(\bof{x})=0} (1-\ep)^{n-d_H(\bof{x},\bof{y})}\cdot \ep^{d_H(\bof{x},\bof{y})},
\sum\limits_{\bof{x}|f(\bof{x})=1} (1-\ep)^{n-d_H(\bof{x},\bof{y})}\cdot \ep^{d_H(\bof{x},\bof{y})}\right\}\\
\nonumber &=&1- \frac{1}{2^n}\sum\limits_{\bof{y}} \max\left\{\sum\limits_{\bof{x}|f(\bof{x})=0} (1-\ep)^{n-d_H(\bof{x},\bof{y})}\cdot \ep^{d_H(\bof{x},\bof{y})},
\sum\limits_{\bof{x}|f(\bof{x})=1} (1-\ep)^{n-d_H(\bof{x},\bof{y})}\cdot \ep^{d_H(\bof{x},\bof{y})}\right\} \\
\nonumber &=& 1-\mathbb{E}_{\bof{y}}[\max(P(f(\bof{X})=0|\bof{Y}=\bof{y}),P(f(\bof{X})=1|\bof{Y}=\bof{y}))]\ .
\end{eqnarray}
\end{small}
However, the last line is exactly equal to $1/2-1/2\cdot c_{f(\bof{X})g(\bof{Y})}$, where $g$ is the maximum likelihood function of $f(\bof{X})$ given $\bof{Y}$. 
\end{proof}

This means that unless Bob is able to create an output bit which is highly correlated with Alice's, the adversary can always obtain knowledge about the key bit. 
However, if Alice just applies the trivial function mapping all outputs to zero, then the correlation between Alice's and Bob's output bit could become $1$, and this bound becomes trivial. 
We will now show that this does not help because in order to obtain a high correlation, Alice and Bob need to apply a biased hash function, and in that case the adversary can obtain high knowledge as well. 

The following theorem, proven by Yang~\cite{Yang07}, shows the trade-off between randomness and correlation of two random bits.
\begin{theorem}[Yang~\cite{Yang07}] \label{th:yang}
Suppose that Alice and Bob share $n$ uniformly random bits with correlation $1-2\ep$. Then the maximal correlation that can be reached if Alice and Bob  both locally  apply a function $f$ (and $g$, respectively) to their $n$ original bits is $1-2\ep(1-4\delta^2)$, where $\delta:=\max(\delta(P_{f(\bof{X})},P_U),\delta(P_{g(\bof{Y})},P_U))$. 
\end{theorem}

Lemma~\ref{lemma:knowledgereduce} shows that if $\delta$ is small, then Eve's knowledge is high. 
We now need to see whether we can lower-bound Eve's knowledge for the case of large $\delta$. For $\delta$ to be large, either $\delta(P_{f(\bof{X})},P_U)$ or $\delta(P_{g(\bof{Y})},P_U)$ needs to be large. Let us first show that if $\delta(P_{f(\bof{X})},P_U)$ is large, then so is Eve's knowledge about $f(\bof{X})$.
But this is easy, as the non-uniformity of the bit $f(\bof{X})$ is automatically also a lower-bound on the non-uniformity of $f(\bof{X})$ as seen from Eve's point of view; i.e., if the key bit is biased, then an adversary has \emph{a priori} information about it. This is stated in Lemma~\ref{lemma:nottoounbalanced}.
\begin{lemma}\label{lemma:nottoounbalanced}
There exists a box partition $w$ such that the non-uniformity of the bit $f(\bof{X})$ is at least $\delta(P_{f(\bof{X})},P_U)$, i.e., $\delta_w^f\geq \delta(P_{f(\bof{X})},P_U)$. 
\end{lemma}
\begin{proof} 
This bound can be obtained by the trivial box partition: 
\begin{eqnarray}
\delta_{no\ partition}^f(\bof{u},\bof{v})&=& \frac{1}{2}\cdot \left|P({f(\bof{X})=0|\bof{uv}})-P({f(\bof{X})=1|\bof{uv}})\right|
=\delta(P_{f(\bof{X})},P_U)\ .
\end{eqnarray}
\end{proof}

We, therefore, found a second bound on the non-uniformity of the bit $f(\bof{X})$ given $w$: $\delta_{no\ partition}^f\geq \delta(P_{f(\bof{X})},P_U)$. 
It remains to exclude the case that $\delta$ is large because $\delta(P_{f(\bof{X})},P_U)$ is small and $\delta(P_{g(\bof{Y})},P_U)$ is large.
\begin{lemma}\label{lemma:wbarbetterthandeltadifference}
There exists a box partition $w$ such that the non-uniformity of the bit $f(\bof{X})$ is at least the absolute value of the difference between $\delta(P_{f(\bof{X})},P_U)$ and $\delta(P_{g(\bof{Y})},P_U)$, i.e., $\delta_{w}^f\geq |\delta(P_{g(\bof{Y})},P_U)-\delta(P_{f(\bof{X})},P_U)|$.
\end{lemma}
\begin{proof}
It is enough to show that the box partition $\bar{w}$ reaches this bound.
Note that when $|\delta(P_{g(\bof{Y})},P_U)-\delta(P_{f(\bof{X})},P_U)|$ is large, then the correlation between the bits $f(\bof{X})$ and $g(\bof{Y})$ must be low:
\begin{eqnarray}
\nonumber c_{f(\bof{X})g(\bof{Y})}&=& 2\cdot P(f(\bof{X})=g(\bof{Y}))-1\\
\nonumber &\leq& 2(1-|\delta(P_{g(\bof{Y})},P_U)-\delta(P_{f(\bof{X})},P_U)|)-1\\
\nonumber &=&1-2|\delta(P_{g(\bof{Y})},P_U)-\delta(P_{f(\bof{X})},P_U)|\ .
\end{eqnarray}
Using Lemma~\ref{lemma:knowledgereduce}, we can connect the correlation with the non-uniformity of $f(\bof{X})$ given~$\bar{w}$:
\begin{eqnarray}
\nonumber |\delta(P_{g(\bof{Y})},P_U)-\delta(P_{f(\bof{X})},P_U)|\leq \frac{1}{2}-\frac{1}{2}c_{f(\bof{X})g(\bof{Y})}= \delta_{\bar{w}}^f\ .
\end{eqnarray}
\end{proof}

Using Lemma~\ref{lemma:nottoounbalanced} and \ref{lemma:wbarbetterthandeltadifference}, we can now connect the non-uniformity of the bit $f(\bof{X})$ with $\delta$:
\begin{lemma}\label{lemma:knowledgebiggerthanhalfdelta}
For every hash function $f$, there exists a box partition $w$ such that $\delta_{w}^f\geq 1/2\cdot \delta$, where $\delta:=\max(\delta(P_{f(\bof{X})},P_U),\delta(P_{g(\bof{Y})},P_U))$ and $g$ is the maximum likelyhood function of $f(\bof{X})$ given $\bof{Y}$. 
\end{lemma}
Note that the box partition $w$ can depend on the choice of hash function $f$ because the adversary can delay the choice of $w$. 
\begin{proof} 
Lemmas~\ref{lemma:nottoounbalanced} and \ref{lemma:wbarbetterthandeltadifference} show that 
\begin{eqnarray}
\nonumber \delta_{no\ partition}^f &\geq & \delta(P_{f(\bof{X})},P_U)\\
\nonumber \delta_{\bar{w}}^f &\geq & \delta(P_{g(\bof{Y})},P_U)-\delta(P_{f(\bof{X})},P_U)\ .
\end{eqnarray}
This implies directly that there exists a suitable box partition (either $\bar{w}$ or the trivial one) such that $\delta_{w}^f\geq 1/2\cdot  \max(\delta(P_{g(\bof{Y})},P_U),\delta(P_{f(\bof{X})},P_U))=1/2\cdot \delta$.
\end{proof}

Now we can put all the previous lemmas together to obtain a general lower bound on the adversary's knowledge. 
\begin{theorem}\label{main}
For every hash function $f$, there exists a box partition $w$ such that the non-uniformity of the bit $f(\bof{X})$ given $w$ is at least $\frac{-1+\sqrt{1+64\ep^2}}{32\ep}$.
\end{theorem}

\begin{figure}[t!]
\centering
\includegraphics[width=7cm]{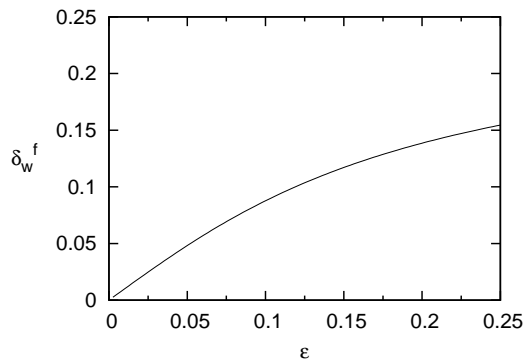}
\caption{\label{fig:non-univ}The lower bound on the non-uniformity of the final bit as function of the error of the boxes $\ep$.}
\end{figure}
\begin{proof}
Lemmas~\ref{lemma:knowledgereduce} and~\ref{lemma:knowledgebiggerthanhalfdelta} show that 
\begin{eqnarray}
\nonumber \delta_{w}^f &\geq & 1/2\cdot \delta\\
\nonumber \delta_{\bar{w}}^f &\geq& 1/2-1/2\cdot c_{f(\bof{X}),g(\bof{Y})}\geq\ep(1-4\delta^2)\ ,
\end{eqnarray}
where $\delta:=\max(\delta(P_{f(\bof{X})},P_U),\delta(P_{g(\bof{Y})},P_U))$. Therefore, $ \delta_{\bar{w}}^f \geq \max( 1/2\cdot \delta,\ep(1-4\delta^2) )$, which takes its lowest value for $1/2\cdot \delta=\ep(1-4\delta^2)$, namely $1/2\cdot\delta=\frac{-1+\sqrt{1+64\ep^2}}{32\ep}$. 
\end{proof}

Note that for small $\ep$, this lower bound actually gives a value of $\delta_{w}^f $ close to $2\ep$; for $\ep$ close to $0.25$, it is still larger than $\ep/2$. We obtain a constant lower bound (see Fig.~\ref{fig:non-univ}) depending only on the error of the individual boxes $\ep$ but not on the number of boxes $n$. This shows that the non-uniformity of the bit $f(\bof{X})$, given $w$, can never become negligible in the number $n$ of boxes, and, therefore, privacy amplification of relativistic cryptography is impossible.

\section{Concluding Remarks}

Cryptographic security can be proven only if certain assumptions are made.
This can be a limitation on the adversary's computing power, memory space,
or accessible information. Another example is quantum cryptography, which
is based on the accuracy and completeness of the quantum-physical description of nature. Although this theory has
been tested by a great number of experiments, it may be attractive to have
an alternative, and to base cryptographic security on the fact that
quantum {\em or\/} relativity theory is correct. An additional advantage --- and more important in practice ---
of such ``non-signaling'' schemes, first
proposed by Barrett,
Hardy, and Kent~\cite{kent}, over traditional quantum cryptography, going back to Bennett and Brassard~\cite{bb84}
as well as Ekert~\cite{ekert}, is that the security is \emph{device-independent}:
Alice and Bob do not have to trust the manufacturer of the devices or,
more precisely, in the fact that they are actually operating on the quantum
systems they are supposed to be. They can derive the security directly
from the correlations in their classical data.

Unfortunately, it appears that such security cannot be achieved this way
unless the physical systems are noiseless and the communication complexity is exponential
in any reasonable security parameter (as it is the case for Barrett {\em et\ al.}'s protocol).
Indeed, we have shown that one of the key ingredients for obtaining
unconditional classical as well as quantum key agreement efficiently, namely privacy
amplification, fails here.
In this light, it may be even more surprising that general quantum privacy amplification 
is possible~\cite{koenigrenner}. 
In particular, note that our impossibility result holds even for the case of 
collective attacks, for which the possibility of device-independent security has been 
shown~\cite{abgs}. An obvious open question is whether privacy amplification could be made possible by enforcing a time-like ordering between the $n$ systems and therefore imposing a non-signaling condition in one direction. Physically, this would be easily realizable by 
measuring several quantum systems one after another.  \\
\\
\textbf{Acknowledgments: } We thank two anonymous referees for their 
valuable comments. This work is supported by the Swiss National Science Foundation. 

\bibliographystyle{hplain}
\bibliography{impossibility_of_ns_pa}

\end{document}